\documentstyle[epsfig]{mn}

\def\be{\begin{equation}}
\def\ee{\end{equation}}
\def\bea{\begin{eqnarray}}
\def\eea{\end{eqnarray}}

\def\etal{{\em et al.}            }

\def\mum{\mu {\rm m}}
\def\kms{{\rm \,km\,s}^{-1}}

\def\kpc{{\rm\,kpc}}

\def\hMpc{\,h^{-1}{\rm Mpc}}

\def\pster{{\rm \,ster}^{-1}}

\def\log{{\rm log}}

\def\Ha{${\rm H}\alpha~$}

\def\degs{{\rm \,deg}^2}

\def\dg{^{\circ}}

\def\Jy{{\rm \,Jy}}
\def\MJy{{\rm \,MJy}}
\def\spose#1{\hbox to 0pt{#1\hss}}
\def\simlt{\mathrel{\spose{\lower 3pt\hbox{$\mathchar"218$}}
     \raise 2.0pt\hbox{$\mathchar"13C$}}}
\def\simgt{\mathrel{\spose{\lower 3pt\hbox{$\mathchar"218$}}
     \raise 2.0pt\hbox{$\mathchar"13E$}}}
\def\({\left(}
\def\){\right)}
\def\[{\left[}
\def\]{\right]}
\def\<{\left\langle}
\def\>{\right\rangle}

\def\AstAst{A\&A~}
\def\AAS{A\&AS~}

\def\ApJ{ApJ~}

\def\ApJS{ApJS~}

\def\ARAA{ARAA~}
\def\MN{MNRAS~}

\def\edcomment#1{\iffalse\marginpar{\raggedright\sl#1\/}\else\relax\fi}

\begin{document}
\title{The PSCz catalogue}
\author[W.Saunders \etal]{W.Saunders$^1$, W.J.Sutherland$^2$, S.J.Maddox$^3$, O.Keeble$^4$, S.J.Oliver$^4$, \cr M.Rowan-Robinson$^4$,  R.G.McMahon$^3$, G.P.Efstathiou$^3$, H.Tadros$^2$, \cr S.D.M.White$^5$, C.S.Frenk$^6$, A. Carrami\~{n}ana$^7$, M.R.S.Hawkins$^1$ \\ $^1$ Institute for Astronomy, Blackford Hill, Edinburgh EH9 3RJ. \\  $^2$ Nuclear and Astrophysics Laboratory, Keble Road, Oxford OX1 3RH. \\ $^3$ Institute of Astronomy, Madingley Road, Cambridge CB3 0HA.\\ $^4$ Blackett Laboratory, Imperial College, Prince Consort Road, London SW7 2BZ. \\ $^5$ MPI-Astrophysik,Karl-Schwarzschild-Strasse 1, Garching bei Munchen, Germany D-85740. \\ $^6$ Department of Physics, University of Durham, DH1 3LE. \\ $7$ Instituto Nacional de Astrof\'{i}sica \'{O}pticta y Electr\'{o}nica, Apartado Postal 51 y 216, 72000, Puebla, Mexico.}
\maketitle

\begin{abstract}

We present the catalogue, mask, redshift data and selection function
for the PSCz survey of 15411 IRAS galaxies across 84\% of the
sky. Most of the IRAS data is taken from the Point Source Catalog, but
this has been supplemented and corrected in various ways to improve
the completeness and uniformity. We quantify the known imperfections
in the catalogue, and we assess the overall uniformity, completeness
and data quality. We find that overall the catalogue is complete and
uniform to within a few percent at high latitudes and 10\% at low
latitudes. Ancillary information, access details, guidelines and
caveats for using the catalogue are given.


\end{abstract}

\section{Introduction}

Data from IRAS (the Infra-Red Astronomical Satellite) allows
unparalleled uniformity, sky coverage and depth for mapping the local
galaxy density field. In 1992, with completion of the QDOT and 1.2 Jy
surveys (Rowan-Robinson \etal 1990a, Lawrence \etal 1999, Strauss
\etal 1990, 1992, Fisher \etal 1995), and with other large redshift
surveys in progress, it became clear that a complete redshift survey
of the IRAS Point Source Catalog (the PSC, Beichman \etal 1984,
henceforth ES) had become feasible.

Our specific targets for the PSCz survey were two-fold: (a) we wanted
to maximise sky coverage in order to predict the gravity field, and
(b) we wanted to obtain the best possible completeness and flux
uniformity within well-defined area and redshift ranges, for
statistical studies of the IRAS galaxy population and its
distribution. The availability of digitised optical information
allowed us to relax the IRAS selection criteria used in the QIGC
(Rowan-Robinson \etal 1990b), and use optical identification as an
essential part of the selection process. This allowed greater sky
coverage, being essentially limited only by requiring that optical
extinctions be small enough to allow complete identifications. The PSC
was used as starting material, because of its superior sky coverage
and treatment of confused and extended sources as compared with the
Faint Source Survey (Moshir \etal 1989). The depth of the survey
($0.6\Jy$) derives from the depth to which the PSC is complete over
most of the sky.

The topology of the survey is analysed and presented by Canavezes \etal (1998); the inferred velocity field by Branchini \etal (1999); the real-space power spectrum and its distortion in redshift-space by Tadros \etal (1999); the redshift-space power spectrum by Sutherland \etal (1999). The direction and convergeance of the dipole has been investigated by Rowan-Robinson \etal (1999), and its implications for cosmological models by Schmoldt \etal (1999). Sharpe \etal (1999) presented a least-action reconstruction of the local velocity field, while an optical/IRAS clustering comparison was presented by Seaborne \etal (1999). Many of these results are summarised in Saunders \etal (1999).

\section{Construction of the Catalogue}

\subsection{Sky coverage}

We aimed to include in the survey all areas of sky with (a) reliable
and complete IRAS data, and (b) optical extinction small enough to
allow reliable galaxy identifications and spectroscopic followup. We
defined a mask of those parts of the sky excluded from the survey; the
final mask was the union of the following areas:

\begin{enumerate} 
\item{Areas failing to get 2 Hours-Confirming coverages (HCONs, ES
III.C.1). In these areas, there is either no data at all or the data
does not allow adequate source confirmation.}
\item{Areas flagged as High Source Density at $12$, $25$ or $60\mum$. HSD at $12$ or $25$ implies an impossibly high stellar density for galaxy identifications. Areas flagged as HSD at $60\mum$ were processed differently in the PSC, with completeness sacrificed for the sake of reliability.}
\item{Areas with $I_{100} > 25 \MJy\pster$. The $100\mum$ intensity values are those of Rowan-Robinson \etal (1991). Above this value, we found the fraction of sources which were identifiable as galaxies dropped dramatically.}
\item{Areas with extinctions $A_B>2^m$, where secure optical identifications become increasingly difficult. Our extinction estimates were based on the $I_{100}$ values above, and incorporated a simple model for the the variation in dust temperature across the Galaxy; they are discussed further in Section 4.3. Except towards the centre and anti-centre, this criterion and the previous one are almost equivalent}
\item{Small patches covering the LMC and SMC, defined by  $I_{100} > 10$ and $5 \MJy\pster$ respectively. In these areas, there are large numbers of HII regions in the clouds themselves with similar optical and IRAS properties to background galaxies, making identifications very uncertain.}
\end{enumerate}

The mask is specified as a list of excluded $1 \degs$ `lune-bins' (ES
Ap.X.1), so these values are necessary averages over $1\degs$. 

The overall area outside the mask is 84\% of the sky. For statistical
studies of the IRAS galaxy population and its distribution, where
uniformity is more important than sky coverage, we made a `high $|b|$'
mask, as above but including all areas with $A_B>1^m$, and leaving
72\% of the sky. In practice, this criterion is almost identical to
$I_{100} > 12.5 \MJy\pster$. Henceforth, when we say `high-latitude',
we simply mean outside this mask. Both masks are shown in Figure 1.

\begin{figure*}
\centerline{\epsfig{figure=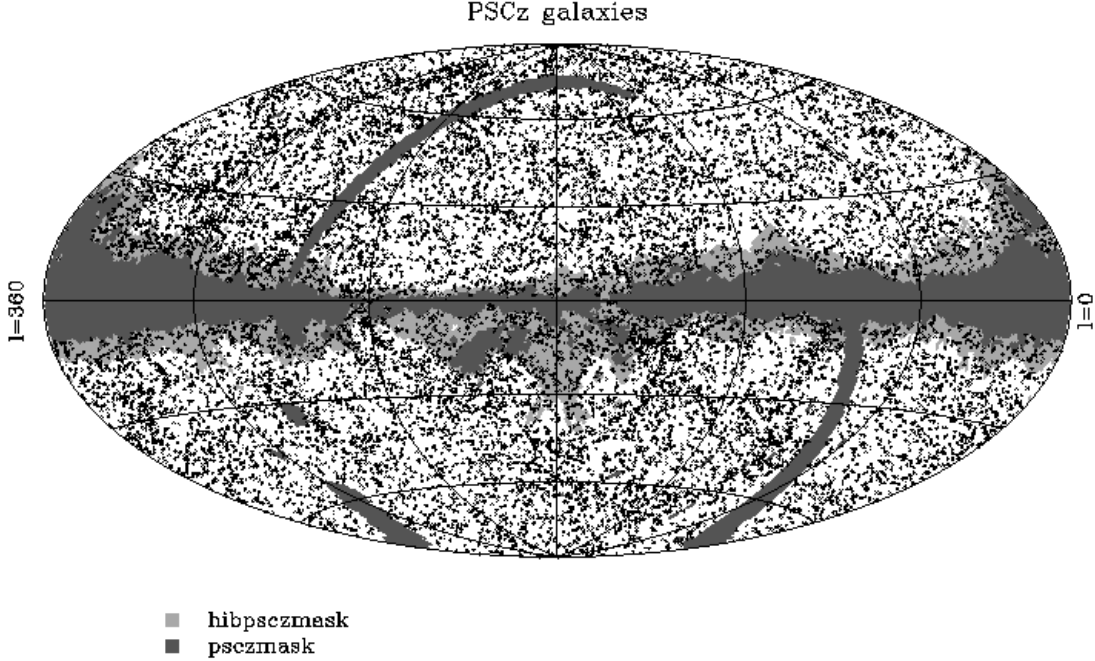,width=12.5cm,angle=-90}}
\caption{15431 PSCz sources identified as possible or definite galaxies, high $|b|$ and default masks in galactic coordinates.}
\end{figure*}

\subsection{PSC selection criteria}

Our aim was to relax the criteria of the QIGC sufficiently to pick up
virtually all galaxies, even at low latitude, purely from their IRAS
properties; simultaneously, we wanted to keep contamination by
Galactic sources to a reasonable level. Our actual colour selection
criteria were almost the same as the QIGC:

\begin{tabular}{lcr} 
$\log_{10}(S_{60}/S_{25})$ &$>$& $-0.3$ \\
$\log_{10}(S_{25}/S_{12})$ &$<$& $1.0$ \\ 
$\log_{10}(S_{100}/S_{25})$&$>$& $-0.3$ \\
$\log_{10}(S_{60}/S_{12})$ &$>$& $0.0$ \\
$\log_{10}(S_{100}/S_{60})$ &$<$& $0.75$ 
\end{tabular}

However, unlike the QIGC, upper limits were used only where they
guaranteed inclusion or exclusion. We did not at this stage exclude
any source solely on the basis of an identification with a Galactic
source. Because the PSCz evolved from the QIGC, there remain in the
catalogue 3 galaxies which actually fail the new selection
criteria. In total, we selected 16422 sources from the PSC.

Many very nearby galaxies have multiple PSC sources, associated with
individual starforming regions. 70 such sources were pruned to leave
for each such galaxy, the single, brightest PSC source. Also at this
stage, Local Group galaxies were excised from the catalogue, and a
separate catalogue of far-infrared properties for Local Group galaxies
made.

\subsection{Extended sources}

All IRAS surveys are bedevilled by the question of how to deal with
galaxies which are multiple or extended with respect to the IRAS
$60\mum$ beam. The size of most $60\mum$ detectors was $1.5'$ in-scan
by $4.75'$ cross-scan. The raw data was taken every $0.5'$, and for
the PSC this was then filtered with an 8-point zero-sum linear filter,
of the form $--++++--$. Hence galaxies with in-scan far-infrared sizes
larger than about $1.5'$, will have their fluxes underestimated in the
PSC. There is a lesser sensitivity to cross-scan diameter, caused by
some of the scans only partially crossing the full width exteneded
sources.

The approach we have settled on is to preferentially use PSC fluxes,
except for sources identified with individual galaxies whose large
diameters are likely to lead to significant flux underestimation in
the PSC. The PSC flux is based on a least chi-squared template fit to
the point-source-filtered data-stream, so the flux is correctly
measured for slightly extended sources, with diameters smaller than
about $1.5'$\footnote[1]{By comparison, the Faint Source Survey
(Moshir \etal 1989) fluxes are peak amplitudes of the
point-source-filtered data, so are much less tolerant of slightly
extended sources.}. As far-infrared diameters are typically half
optical, (e.g. Rice \etal 1988), and optical $D_{25}$ diameters are
typically several times the FWHM for spirals, we can expect that
isolated galaxies must have extinction-corrected diameters larger than
several arcmin for their fluxes to be badly underestimated by the PSC.

Fluxes for extended sources can be derived using the ADDSCAN (or
SCANPI) software provided at IPAC, which coadds all detector scans
passing over a given position. We have tested how the ratio of
PSC-to-extended flux depends on $D_{25}$, and, as expected, we find
that the ratio is almost constant for galaxies up to $2.5'$ (Figure
2), where the addscan flux is typically 10\% larger. However, we also
find that addscan fluxes are systematically 5\% larger than PSC
fluxes, even for much smaller galaxies. This discrepancy is not due to
any difference in calibration (we have tested this by extracting a
point-source-filtered flux from the addscan data and substituting this
for the PSC flux, but the discrepancy in Figure 2 remains). The reason
is the large number of multiple and/or interacting galaxies seen by
IRAS, for which the addscan picks up the entire combined flux. Using
such combined fluxes throughout is not a consistent approach, since
nearby interacting galaxies will always be resolved into two or more
sources, while more distant ones will not be\footnote[2]{In {\em any}
survey with unresolved sources, a consistent approach to multiple
sources is not possible, except by artificially degrading the
resolution for each source to make it the same in physical units as
for the most distant.}. The effect of the PSC filter is more subtle: a
close pair of galaxies unresolved by the IRAS beam will in general
have unequal fluxes, and in a flux limited survey, at least one will
be below the flux limit. In such a situation, the zero-sum linear PSC
filter still overestimates the flux of the brightest source on average
(because of clustering), but by much less than the addscan.

Very large galaxies will have fluxes underestimated by the addscan,
because their cross-scan extent is large compared with the detector
size. Fluxes for galaxies with optical diameters $D_{25}>8'$ have been
taken from Rice \etal (1988), who made coadded maps for each such
galaxy. In principal, galaxies somewhat smaller than this may have
underestimated fluxes from the addscans. However, we find that, for
the 19 PSCz galaxies in Rice \etal with quoted diameters $8'-10'$, the
flux differance between addscan and Rice \etal fluxes is
$\log_{10}(S_{60A}/S_{60R})=0.044\pm0.034$, that is the addscan fluxes are
slightly but not significantly larger; so we do not expect smaller
galaxies to have underestimated addscan fluxes.

\begin{figure}
\centerline{\epsfig{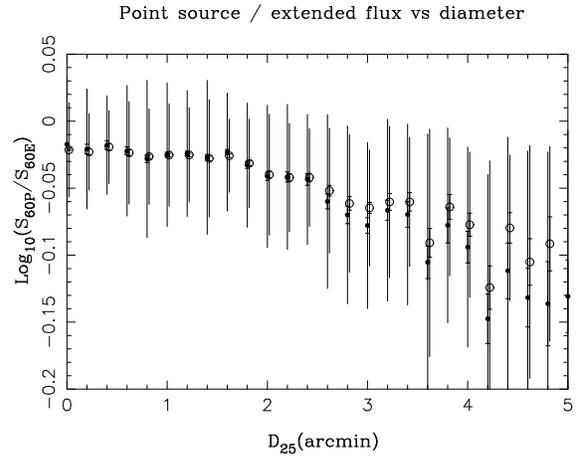}}
\caption{Ratio of extended flux to PSC flux (filled circles) or point-source-filtered addscan flux (open circles), as a function of optical diameter. The open circles are displaced slightly to the right for clarity. The inner error bars are the uncertainty on the ratio, the outer error bars are the population scatter. Where $D_{25}$ is not known, it is set to 0.}
\end{figure}

Bearing these points in mind, we proceeded as follows: we made a
catalogue, with the same sky coverage as the PSCz, of
optically-selected galaxies from the LEDA database (Paturel \etal
1989) with extinction-corrected $D_{25}$ diameters larger than
$2.25'$, where the extinctions were estimated as per Section 4.3, and
the consequent corrections to the diameters are as given by Cameron
(1990). In Saunders \etal (1995), it was argued that LEDA is
reasonably complete to this limit. For the largest sources, we used
the positions and fluxes of Rice \etal (1988). IPAC kindly addscanned
all the remainder for us to provide coadded data. We then used
software supplied by Amos Yahil to extract positions, fluxes and
diameters from this data, on the assumption that the galaxies have
exponential profiles. Where these addscan fluxes are used in the
catalogue, they have been arbitrarily decreased by 10\%, to bring them
statistically into line with the PSC fluxes at the $2.25'$
switchover. We ended up with 1402 sources associated with large
galaxies entering the catalogue, and 1290 PSC sources associated with
large galaxies flagged for deletion. The latter are kept in the
catalogue, so that a purely IRAS-derived catalogue can be extracted if
wanted.

\subsection{Other problem sources}

In the QIGC, sources flagged as associated with confirmed extended
sources, or with poor Correlation Coefficient, or flagged as confused,
were addscanned. The above prescription should have dealt with the
extended sources; the only other reasons for a galaxy source to have a
poor CC are that (a) it is multiple, and as argued above we prefered
PSC fluxes in these cases, or (b) it has low S/N, in which case the
PSC is an unbiased flux estimator, while addscanning risks
noise-dependent biases. The confusion flag is set very conservatively
(i.e. most sources flagged as confused have sensible PSC fluxes) and
in any case the PSC in general deals better with confusion than the
addscans, in both cross-scan and in-scan directions. For these
reasons, and to avoid introducing any latitude-dependent biases into
the catalogue, we opted to use PSC fluxes throughout for sources not
identified with optical galaxies as above. Addscan fluxes are included
in the catalogue for all sources for those wishing to experiment with
them.

\subsection{Supplementary sources}

For a source to be accepted into the PSC, it required successful
Hours-Confirmation on two separate HCONS (ES V.D.6). Thus at low S/N,
areas of the sky with only 2 HCONS are inevitably less complete than
those with 3 or more; the estimated completeness at $0.6-0.65\Jy$ in
2HCON areas is only 82\% (ES XII.A.4). To improve this completeness,
we supplemented the catalogue with 1HCON sources satisfying our colour
criteria in the Point Source Catalog Reject File (ES VII.E.1), where
there is a corresponding source in the Faint Source Survey. We
demanded that PSC and FSS sources be within each other's $2 \sigma$
error ellipse, and that the fluxes agree to within a factor of
$1.5$. This revealed many sources in the Reject File where two
individual HCON detections had failed to be merged in the PSC
processing (ES XII.A.3), as well as sources which failed at least one
HCON for whatever reason. New sources were created or merged with
existing ones, and the Flux Overestimation Parameters (ES XII.A.1)
assigned or recalculated accordingly. As well as these 1HCON sources,
we also searched in the Reject File for additional sources with flux
quality flags 1122 and 1121 \footnote[3]{1=upper limit only,
2=moderate quality detection, 3=good detection, in each of the four
bands}; neither of which category made the PSC. Finally, we looked for
sources in the PSC itself with flux quality flags 1113, but
correlation coefficient at $60\mum$ equal to A,B or C indicating a
meaningful detection. Altogether we found an additional 323 galaxies,
mostly in 2HCON regions, and made 143 deletions as a result of merging
individual HCON detections.

\subsection{Optical identifications}

Optical material for virtually all sources was obtained from COSMOS or
APM scans, including new APM scans taken of 150 low-latitude POSS-I E
plates. In general, we used red plates at $|b|<10\dg$ and blue
otherwise. The actual categories of optical material are:\\
1. Uncalibrated POSS-O plates scanned with APM. Here the magnitudes
are typically good to $0.5^m$ for faint ($B\sim19^m$) galaxies.\\
2. SRC J plates scanned with APM. These plates were scanned and
matched for the APM survey, and have $b_J$ magnitudes typically good
to $0.25^m$ for faint galaxies.\\ 3,4,5. Uncalibrated SRC J or EJ
plates scanned with COSMOS, giving $b_J$ accurate to about $0.5^m$. \\
6. SRC SR plates scanned with COSMOS. The quoted magnitudes are
r-magnitudes, good to $0.5^m$.\\ 7. POSS-E plates scanned with
APM. For these plates, the standard POSS-O calibration was
assumed. Since the O plates are about 1 magnitude deeper than the E
plates, while $B-R \sim 1$ for IRAS galaxies, this gives a reasonable
approximation to the $B$ magnitude. When there is extinction, this
procedure gives systematically too bright a magnitude by about $0.43
A_B$ (assuming Mathis 1990). The scatter is still dominated by the
$0.5^m$ zero-point error.

At bright magnitudes, photographic photometry for galaxies becomes
increasingly uncertain because of non-linearity and saturation in the
emulsion.

The image parameters from COSMOS and APM data were also used to get
arcsecond positions and offsets from nearby stars, and to make
`cartoon' representations of the $4'\times 4'$ field centred on each
IRAS source. The identifications were made using the likelihood
methods of Sutherland and Saunders (1992). In general, and always at
low latitudes, these cartoons were supplemented by grayscale images
from the Digital Sky Survey and/or inspection of copy plates, and any
change in best identification noted.

At this stage, non-galaxies were weeded out by a combination of
optical appearance, IRAS colours and addscan profiles, VLA 20cm maps
from the NVSS survey (Condon \etal 1998), millimetre data from
Wouterlout and Brand (1989), SIMBAD and other literature data. Several
hundred low-latitude sources were also imaged at $K'$, as part of the
extension of the PSCz to lower latitudes (Saunders \etal 1999). We
found a total of 1376 confirmed non-galaxies. We are left with 15332
confirmed galaxies, and a further 79 sources where no optical
identification is known but there is no clear Galactic identification
either. The distribution of identified galaxies and the mask are shown
in Figure 1.

The Galactic sources are dominated by infrared cirrus (774
sources). However, there are also 88 planetary nebulae, 140
emission-line stars, 24 sources identified with Galactic HII regions,
212 sources identified with bright stars or reflection nebulae, and
138 sources associated with YSO's.

\section{The redshift survey}

An essential part of the project was maintenance of a large database
of redshift information from the literature, from databases such as
the NED and LEDA (the NASA and Lyons Extragalactic Databases) and
Huchra's ZCAT, and also work in progress for other surveys. Redshifts
were accepted on the basis that their claimed error was better than
any other available for that source, including our own.

Of the 15,411 galaxies in the sample, about 8,000 had known redshift
at the inception of the project, with about another 2,000 expected to
be observed as part of ongoing projects such as the CfA2 and SSRS
surveys. The project was fortunate enough to be allocated 6 weeks of
INT+FOS time, 1 week INT+IDS, 6 nights AAT+FORS time, 18 nights on the
CTIO 1.5m, two weeks at the INAOE 2.1m and 120 hours at Nan\c{c}ay
radiotelescope, over a total of 4 years. 4600 redshifts were obtained
in this time. For the INT+IDS and CTIO spectra, redshift determination
was made from the average observed wavelengths of the \Ha, NII and SII
features. The rest of the data was taken with low dispersion
spectrographs with resolutions around 15-20 Angstroms, and the \Ha/NII
lines are blended together, as are the two SII lines. We modelled each
continuum-subracted spectrum as a linear combination of \Ha, NII and
SII features, with redshift as a free parameter. The model giving the
smallest $\chi^2$ versus the data gave the redshift and its error,
also the various linestrengths and a goodness of fit. Spectra with
poor $\chi^2$, large redshift uncertainty or unphysical line ratios
were checked by hand and where necessary refitted with different FWHM
or initial guesses. Wavelength calibration included cross-correlation
of each sky spectrum with a well calibrated template. The final
derived errors average $120 \kms$, as compared with $300 \kms$ for the
QDOT survey. Further details are presented in Keeble (1996), and
Oliver \etal (1999).

By construction, we have minimised the overlap with other redshift
surveys, but there are to date 448 galaxies for which we have both our
own measurement, and higher resolution measurements by other workers
(Figure 3). There are 31 sources where the velocity difference is more
than $3 \sigma$ (errors combined in quadrature), if these are clipped
out, the remaining 417 sources give an average offset $V_{PSCz} -
V_{lit} = -6.1 \pm 6.1 \kms$, with a scatter of $124 \kms$ and a
reduced $\chi^2_\nu = 0.89$ . The scatter includes the contribution
from the non-PSCz redshift, and suggests that our quoted errors are
good estimates of the genuine external error. Of the 31 discrepant
redshifts, 5 involve HI observations with unrealistically small quoted
errors and with absolute differences less than $250\kms$, a further 11
are between 100 and 600 $\kms$ and $3-5 \sigma$, and suggest a
non-Gaussian error distribution, and 15 are greater than $700 \kms$
and $5 \sigma$ and must represent different galaxies or incorrect line
identification or calibration. We thus have 15 redshifts seriously in
error out of a total of 896 recent measurements (both ours and other
workers) giving an error rate of 1.7\%.

\begin{figure}
\centerline{\epsfig{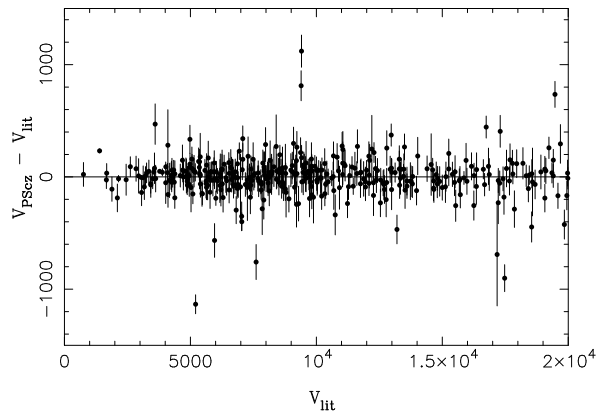}}
\caption{PSCz versus literature redshifts. Error bars show combined errors.}
\end{figure}

We did not pursue very faint ($b_J > 19.5^m$) galaxies, on the basis
that (a) they were certain to be at distances too large to be included
for either statistical or dynamical studies, and (b) very large
amounts of telescope time would be needed to achieve useful
completeness for these galaxies. There are 438 sources without known
redshift with a clear or probable faint optical galaxy identification
(with $b_J > 19.5^m$), and a further 127 with no obvious optical
identification but no secure identification as a Galactic source
either. At time of writing we are still lacking redshifts for 189
brighter ($b_J < 19.5^m$) galaxies. The sky distribution of these
categories is shown in Figure 4. Redshifts are available for 14677
sources.

\begin{figure*}
\centerline{\epsfig{figure=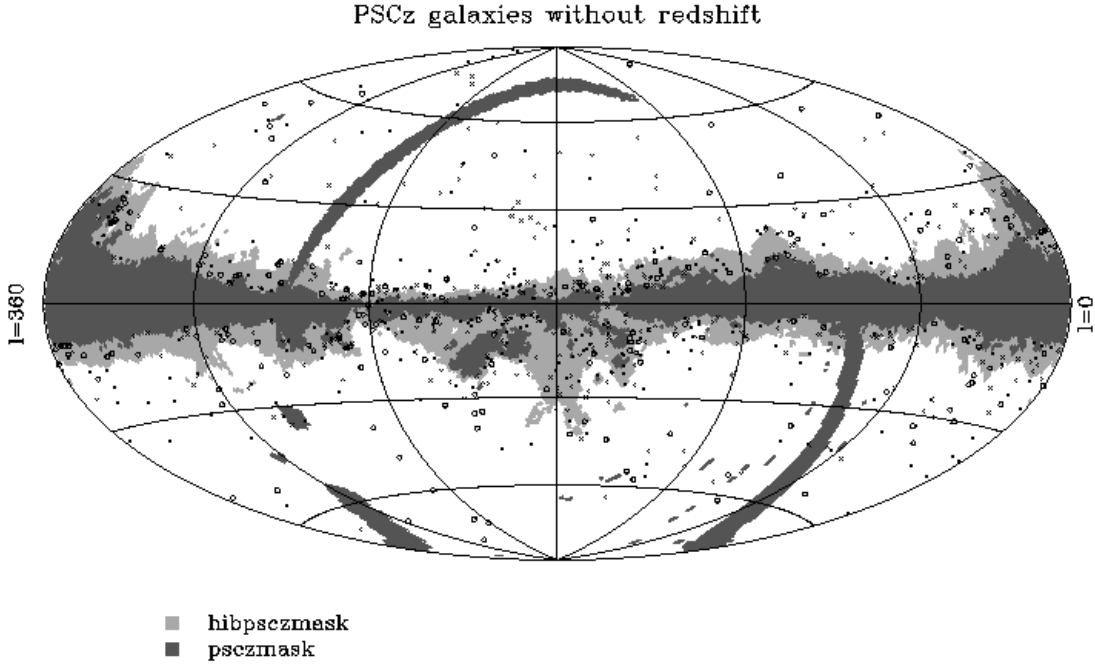,width=12.5cm,angle=-90}}
\caption{Sky distribution of PSCz sources without redshift, with galaxy identifications $b_J<19.5^m$ (open circles, 192 sources) and fainter than this or with no identification (crosses, 542 sources).}
\end{figure*}

\subsection{n(z) and selection function}

The $n(z)$ distribution is shown in Figure 5. The source density
amounts to 1460 gals $\pster$ at high latitudes and the median redshift is
$8500 \kms$.  To account for the effect of the flux limit on the
observed number density of galaxies as a function of distance, we need
to know the {\em selection function} $\psi(r)$, here defined as the
expected number density of galaxies in the survey as a function of
distance in the absence of clustering. We have derived this both
non-parametrically and parametrically using the methods of Mann,
Saunders and Taylor (1996). This method is almost completely
insensitive to the assumed cosmology, in the sense that the derived
expected $n(z)$ is invariant. The resulting selection function can be
transformed to other cosmologies or definitions of distance simply by
mapping the volume element or distance, keeping $n(z)dz$ invariant.

For simplicity, and to allow comparison with simulations, we assume
for derivation purposes a Euclidean Universe without relativistic
effects, and with distance $r$ equal to $V/100h\kms$. We derive the
result both non-parametrically, and parametrically using the double
power-law form

\be  \psi(r) = \psi_* {\({r \over r_*}\)}^{1-\alpha} {\[{1+{\(r\over 
r_*\)}^\gamma}\]}^{- \(\beta \over \gamma\)} \ee

\parindent=0pt which very well describes the non-parametric
results. The parameters $\psi$, $\alpha$, $r_*$, $\gamma$ and $\beta$
respectively describe the normalisation, the nearby slope, the break
distance in $\hMpc$, its sharpness and the additional slope beyond it.

Using the high-latitude PSCz to derive the selection function, and
correcting redshifts only for our motion with respect to the centroid
of the local group, $V=V_{hel}+300 {\rm sin} l {\rm cos} b$, we obtain

 $\psi_*= 0.0077, \alpha =  1.82, r_*= 86.4, \gamma=  1.56, \beta=  4.43$

Both non-parametric and parametric results are shown in Figure 6. The
uncertainty in the selection function is less than 5\% for distances
$30-200\hMpc$, and 10\% for $10-300\hMpc$, although the selection
function drops by 4 decades over this range.

In Figure 6, we also show the non-parametric selection function
derived by the same method, for the QDOT and $1.2\Jy$ surveys.

\begin{figure}
\centerline{\epsfig{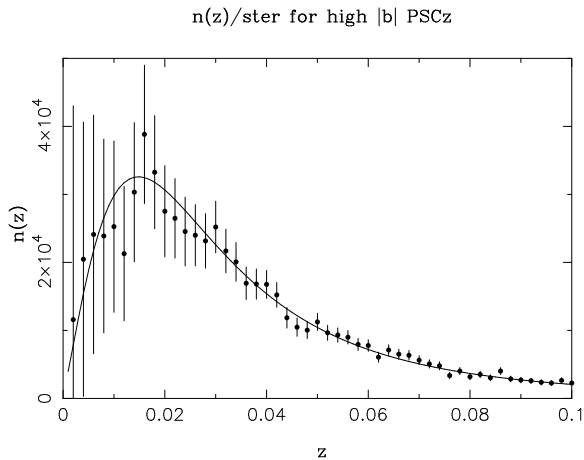}}
\caption[]{$n(z)$ distribution for high latitude PSCz survey. Error bars are $J_3$-weighted. The line is the prediction from the selection function.}
\end{figure}
\begin{figure}
\centerline{\epsfig{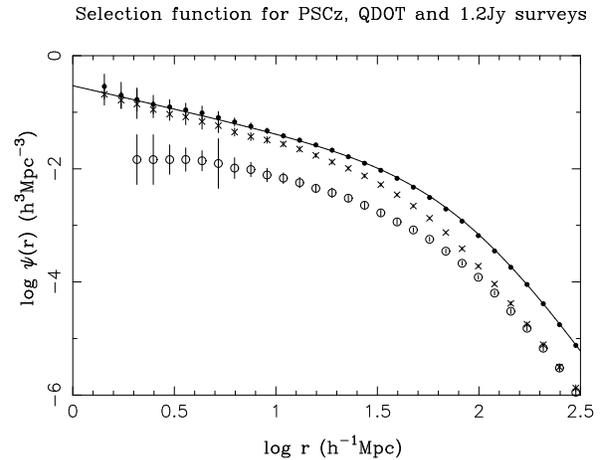}}
\caption[]{Parametric and non-parametric selection functions for high latitude PSCz survey (line and filled circles). Also shown are the QDOT (open circles) and 1.2Jy (crosses) non-parametric selection functions.}
\end{figure}

\section{Reliability, completeness, uniformity, flux accuracy}

The utility of the PSCz for cosmological investigation depends on its
reliability, completeness and uniformity.

\subsection{Reliability}
The major sources of unreliability in the catalogue are (a) incorrect
identifications with galaxies nearby in angular position, and (b)
incorrect identification of spectral features. Later cross-referencing
used increasingly sophisticated methods based on Sutherland and
Saunders (1992), but many older identifications depend on simple $2'$
proximity. The most recent cross-correlation with ZCAT provided 1975
updated velocities, of which 84 were altered by more than $500\kms$
and 50 by more than $1000\kms$, suggesting an error rate of 2\%, in
agreement with the analysis of Section 3.

The identifications made for our own redshift followup were done much
more carefully than those taken from the literature, using a
likelihood approach including full knowledge of the error ellipse,
optical galaxy source counts etc, and checked by eye from a greyscale
image. The number of ambiguous identifications is negligible, though
the PSC source is often confused between two galaxies separated by a
few arcmin in the cross-scan direction. Where there was real
ambiguity, we took redshifts for both galaxies and used the one with
larger \Ha flux.  It is also possible that Galactic sources were
occasionally identified as background galaxies, but we believe this to
be very rare.

\subsection{Completeness}

Incompleteness enters into the catalogue for many reasons:

1) There will be IRAS incompleteness where sources with true $60\mum$
   flux greater than $0.6\Jy$ fail to appear in the catalogue. The
   largest source of this incompleteness is the areas of sky with only
   2HCONs, where the PSC incompleteness is estimated as 20\%
   (differential) and 5\% (cumulative) at $0.6\Jy$ (ES XII.A.4)
   . Based on source counts, we estimate that our recovery procedure
   (Section 2.5) for these sources has reduced the overall
   incompleteness in these areas from 5\% to 1.5\%; Figure 7 shows the
   source counts for high-latitude 2HCON sky. Our recovery procedure
   was impossible for $|b|<10\dg$, so lower-latitude 2HCON sky (2\% of
   the catalogue area) retains the PSC incompleteness.

2) The PSC is confusion limited in the Plane. However, by
   construction, the PSCz does not cover any areas with High Source
   Density at $60\mum$ (as defined in ES V.H.6), and the mask defined
   at $100\mum$ effectively limits the number of confusing
   sources. Also, at low latitudes, there are known problems in the
   PSC with noise lagging. However, the source counts shown in Figure
   8 show low-latitude incompleteness to be small down to $0.6\Jy$.

3) Some galaxies are excluded by our colour criteria. Cool, nearby
   galaxies may fail the $100/60\mum$ condition, but will normally be
   included separately as extended sources. From the comparison with
   the $1.2\Jy$ survey, we estimate that about 50 galaxies from the
   PSC have been excluded (see Section 5).

4) No attempt was made to systematically obtain redshifts for galaxies
   with $b_J > 19.5^m$. At high latitudes, the work of e.g. Clements
   \etal (1996) shows that these will in general be further than
   $z>0.1$. At lower latitudes, incompleteness cuts in at lower
   redshifts. It was originally hoped that the PSCz would be
   everywhere complete to $z=0.05$, but it is clear from the 3D
   density distributions that there is significant (10\% or more )
   incompleteness down to z=0.04 towards the anti-centre. The reasons
   for this are discussed in the next subsection.

Patchy extinction can lead to higher local extinction than our values,
which are averages over $1\degs$ bins. We have obtained $K'$ images of
most low-latitude sources without obvious galaxy or Galactic
identification; many faint galaxies are revealed, but only a handful
of nearby ones. Spectroscopy of the missing low-latitude galaxies is
continuing as part of the Behind The Plane extension to the PSCz
survey; for the time being, we estimate that good completeness has
been achieved for $z < 0.1 /10^{(0.2(A_B+A_B^2/10))}$. The
incompleteness is strongly concentrated towards the anticentre and at
low latitudes; for $|b|>10 \dg$, the survey is estimated to be useably
complete to $z=0.05$ everywhere.

The number of galaxies with $b_J < 19.5^m$ for which redshifts are
still unknown is 192, or 1.2\% of the sample. A further 85 have only
marginal redshift determinations and some of these will be
incorrect. Both unknown and marginal redshifts are well distributed
round the sky, but with some preference for lower latitudes.

\subsection{Extinction maps}

Our extinction maps started from the $I_{100}$ intensity maps of
Rowan-Robinson \etal (1991), binned into lune bins, with the
$I_{100}/A_V$ ratio as given by Boulanger and Perault (1988), and
$A_B/A_V$ ratio as given by Mathis (1990). Because the dust
temperature declines with galactic radius, this procedure
over-estimates extinctions towards the galactic centre, and
under-estimates them towards the anti-centre. We attempted to correct
for this dependence of dust temperature on position by devising a
simple model for the dust and starlight in our galaxy. We assumed a
doubly exponential, optically thin distribution of dust and stars,
with standard IAU scalelengths for the solar radius and stellar
distribution ($r_0=8.5 \kpc, r_{sc}=3.5 \kpc, z_{sc}=0.35 \kpc$) and
an assumed the dust to have the same radial and half the vertical
scalelength. The dust properties were assumed to be uniform, and we
assumed a dust temperature of $20K$ at the solar radius, and
proportional to the one fifth power of the local stellar density
elsewhere. We then found, for each position on the celestial sphere,
the ratio of $100\mum$ emission to column density of dust, and
normalised this ratio by the Boulanger and Perault value for the NGP.

Subsequent to our definition of the PSCz catalogue and mask, Schlegel
\etal (1998) used the IRAS ISSA and COBE DIRBE data to make beautiful
high resolution maps of the dust emission and extinction in our
Galaxy. Comparison of our own maps and those of Schlegel \etal show
that (a) our temperature corrections across the galaxy are too small,
and (b) there are several areas where cooler (16K) dust extends to
$|b|\sim 30 \dg$.

Overall, our extinction estimates may be in error by a factor 1.5-2,
in the sense of being too low towards the anticentre and too high
towards the galactic centre. This has two principal effects. (a)
Towards the anticentre, galaxies may fall below the $b_J=19.5^m$ limit
because of extinction. A subsequent program of $K'$-imaging has
revealed a handful of extra nearby galaxies, and a significant number
at redshifts 0.05 and greater. (b) The definition of the
optically-selected catalogue in Section 2.3 depends on the extinction
corrections. We will have selected too many optical galaxies towards
the centre and too few towards the anticentre. However, our matching
of addscan and PSC fluxes was designed to be robust to exactly this
sort of error. To test for any effect, we have compared the simple,
number-weighted dipole of the surface distribution of PSCz galaxies,
(a) using the normal catalogue and (b) using purely PSC-derived fluxes
as described in section 2.3. The dipole changes by $1\dg$ in direction
and 2.5\% in amplitude when we do this, showing that any bias caused
by incorrect extinctions is negligible.

\subsection{Flux accuracy and uniformity}

The error quoted for PSC $60\mum$ fluxes for genuine point sources is
just 11\%(ES VII.D.2). We are more concerned with any non-fractional
random error component, since this may lead to Malmquist-type
biases. The analysis in Lawrence \etal (1999), based on the
$12/60\mum$ colours of bright stars, finds an absolute error of $0.059
\pm 0.007 \Jy$, in addition to a fractional error of 10\%. Stars are
better point sources than galaxies, so this may underestimate the
absolute error for our sample. We have made an independent estimate,
by investigating the scatter between our PSC and addscan fluxes. Of
course PSC and addscan fluxes start from the same raw data, but the
processing and background estimation are entirely different, while the
actual photon noise is negligible. The absolute component of the
scatter is $0.06-0.07\Jy$, of which an estimated 20\% comes from the
error in the addscan flux - confirming that the Lawrence \etal value
is a reasonable estimate of the PSC absolute error component.

This error estimate leads to Malmquist biases in the source densities
of order 5-6\% (differential, at the flux limit) and 2-3\%
(cumulative) (Murdoch, Crawford and Jauncey 1973). 2HCON sky may have
biases as large as 10\% (differential) and 4\% (cumulative), so the
non-uniformity introduced into the catalogue should be no worse than
5\% (differential) and 2\% (cumulative). This is borne out by the
source counts in Figures 7 and 8, but note that in any noise-limited
catalogue such as the PSC, where the noise varies across the
catalogue, there will always be a regime where Malmquist biases are
masked by incompleteness.

Lawrence \etal (1999) find evidence for slight non-linearity in the
PSC flux scale, but this should not affect any analysis except
evolutionary studies. The effect on evolution was considered by
Saunders \etal (1997).

Non-uniformity can also come about as a result of simple changes in
the absolute flux scale. A small fractional error in the calibration
will on average lead to an error half again as large in the source
density. There are three obvious reasons for flux scale variations:

1) The absolute calibration of the third 3HCON was revised by a few
   percent after the release of the PSC (and FSS). The effect of this
   revision on PSC fluxes would be to change those in the 75\% of the
   sky covered by 3HCONS by 1\%.

2) Whenever the satellite entered the South Atlantic Anomaly,
   radiation hits altered the sensitivity of the detectors, and data
   taken during such times were discarded (ES III.C.4). However, this
   leaves the possibility of Malmquist effects, and also data taken
   near the boundaries of the SAA potentially suffers residual
   effects. We have investigated this by checking the source counts
   for declinations $-40\dg < \delta < 10\dg$, where data taken is
   most likely to be affected. We see no evidence for any variation
   (Figure 7) above the few \% level in source counts.

3) Whenever the satellite crossed the Galactic Plane, or other very
   bright sources, the detectors suffered from hysteresis. This effect
   was investigated by Strauss \etal (1990). They found that the
   likely error is typically less than 1\% and always less than
   2.2\%. This is confirmed by the constancy, to within a few percent,
   of our galaxy source counts for identified galaxies in the PSCz as
   a function of $I_{100}$.

Overall, differential source densities across the sky due to
incompleteness, Malmquist effects and sensitivity variations are not
believed to be greater than a few percent anywhere at high latitudes
for $z<0.1$. Tadros \etal (1999) found an upper limit to the rms
amplitude of large scale, high latitude spherical harmonic components
to the density field of the PSCz of 3\%. Since this is close to the
expected variations due to clustering, the variations due to
non-uniformity in the catalogue must be smaller than this. At lower
latitudes, variations are estimated to be no greater than 10\% for
$z<0.05$.

\begin{figure}
\centerline{\epsfig{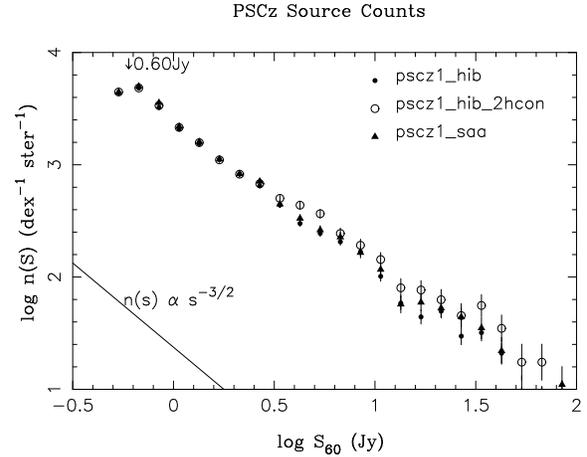}}
\caption[]{Source counts for all high latitude sky (filled circles), high latitude 2HCON sky (open circles), and area with potential SAA problems (triangles).}
\end{figure}

\begin{figure}
\centerline{\epsfig{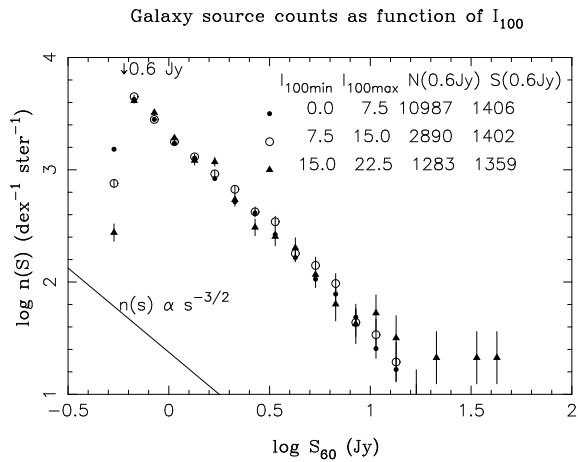}}
\caption[]{Source counts for identified PSCz galaxies as a function of $100 \mum$ background, $I_{100} < 7.5 \MJy\pster$ (filled circles), 7.5-15 (open circles) and 15-22.5 (triangles).}
\end{figure}

\section{Comparison with the $1.2\Jy$ survey}

The $1.2\Jy$ survey of 5500 IRAS galaxies (Fisher \etal 1995) used
looser colour criteria than the PSCz, and as such acts as a valuable
check on the efficacy of our selection procedure from the PSC. We find
that the $1.2\Jy$ survey contains 11 galaxies that have been excluded
by our selection criteria. It contains a further 5 galaxies that
satisfy the PSCz criteria, but are not included due to programming and
editing errors, either during construction of the QIGC survey or its
various extensions and supplements to form the PSCz. Extrapolating
these numbers to lower fluxes, we can expect that about 50 PSC
galaxies are missing altogether from the catalogue.

Conversely, the conservative selection criteria for the $1.2\Jy$
survey led to much greater levels of contamination by cirrus and other
Galactic sources than in the PSCz. In the $1.2\Jy$ survey, these were
eliminated by visual inspection of sky survey plates; inevitably real
galaxies occasionally got thrown out by mistake, especially at low
latitudes. We have found, within the PSCz area, 110 galaxies which are
misclassified as Galactic in the $1.2\Jy$ survey, and we have obtained
redshifts for 67 of them. Most of the remainder are fainter than our
$b_J=19.5^m$ cutoff. There are also known to be to date a further 117
galaxies classified as Galactic in the $1.2\Jy$ survey outside the
PSCz area. These were found as part of the ongoing `Behind the Plane'
extension of the PSCz to lower latitudes described in more detail in
Saunders \etal (1999).

\section{Ancillary information}

Along with the PSC data, the catalogue also contains the following
information: POSS/SRC plate and position on that plate; the
RA,$\delta$, offset, diameters and magnitude of the best match from
the digitised sky survey plates; name, magnitude and diameters from
UGC/ESO/MCG, PGC name, de Vaucouleurs type and HI widths where
available; most accurate available redshift and our own redshift
measurement; classification as galaxy/cirrus/etc; estimated $I_{100}$
and extinction; addscan flux and width when treated as an extended
source, and point source filtered addscan flux.

\section{Accessing the catalogue}

The data is available from the CDS catalogue service (http://cdsweb.u-strasbg.fr/Cats.html). Full and short versions of the catalogue, maskfiles, description files, format statements and notes, are also available via the PSCz web site \\
(http://www-astro.physics.ox.ac.uk/$\sim$wjs/pscz.html), or by anonymous ftp from ftp://ftp-astro.physics.ox.ac.uk/pub/users/wjs/pscz/ .

\section{Acknowledgements}

The PSC-z survey has only been possible because of the generous
assistance from many people in the astronomical community. We are
particularly grateful to John Huchra, Tony Fairall, Karl Fisher,
Michael Strauss, Marc Davis, Raj Visvanathan, Luis DaCosta, Riccardo
Giovanelli, Nanyao Lu, Carmen Pantoja, Tadafumi Takata, Kouichiro
Nakanishi, Toru Yamada, Tim Conrow, Delphine Hardin, Mick Bridgeland,
Renee Kraan-Kortweg, Amos Yahil, Ron Beck, Esperanza Carrasco, Pierre
Chamaraux, Lucie Bottinelli, Gary Wegner, Roger Clowes and Brent
Tully, for the provision of redshifts prior to publication or other
software or data. We are also grateful to the staff at IPAC and the
INT, AAT, CTIO and INOAE telescopes. We have made very extensive use
of the ZCAT, NED, LEDA, Simbad, VLA NVSS and STSCI DSS databases.

\end{document}